\documentclass[12pt]{article}
\usepackage{graphicx}
\usepackage{amsfonts}
\usepackage{amssymb,amsmath}
\usepackage{latexsym}
\usepackage{color}
\usepackage{cite}
\input{colordvi.tex}

\begin{document}

\begin{titlepage}

\begin{flushright}
ICRR-Report-552-2009-14
\end{flushright}

\begin{center}

{\Large \bf  Neutrino mass from cosmology: \vspace{0.2cm} \\
Impact of high-accuracy measurement of  \\ the Hubble constant}

\vskip .45in

{\large
Toyokazu Sekiguchi$^1$, 
Kazuhide Ichikawa$^2$, 
Tomo Takahashi$^3$\\
and
Lincoln Greenhill$^4$
}

\vskip .45in

{\em
$^1$Institute for Cosmic Ray Research, University of Tokyo,
Kashiwa 277-8582, Japan  \vspace{0.2cm} \\
$^2$Department of Micro Engineering, Kyoto University, 
Kyoto 606-8501, Japan \vspace{0.2cm} \\
$^3$Department of Physics, Saga University, Saga 840-8502, Japan   \vspace{0.2cm} \\
$^4$ Harvard-Smithsonian Center for Astrophysics, 
60 Garden Street, \\Cambridge,  MA 02138, USA \vspace{0.2cm}\\
}

\end{center}

\vskip .4in

\begin{abstract}

  Non-zero neutrino mass would affect the evolution of the Universe in
  observable ways, and a strong constraint on the mass can be achieved
  using combinations of cosmological data sets. We focus on the power
  spectrum of cosmic microwave background (CMB) anisotropies, the
  Hubble constant $H_0$, and the
  length scale for baryon acoustic oscillations (BAO) to investigate
  the constraint on the neutrino mass, $m_\nu$. We analyze data from
  multiple existing CMB studies (WMAP5, ACBAR, CBI, BOOMERANG, and
  QUAD), recent measurement of $H_0$ (SHOES), with about two times
  lower uncertainty (5\,\%) than previous estimates, and recent
  treatments of BAO from the Sloan Digital Sky Survey (SDSS).  We
  obtained an upper limit of $m_\nu < 0.2~\mbox{eV}$ (95 \% C.L.), for
  a flat $\Lambda$CDM model. This is a 40\,\% reduction in the limit
  derived from previous $H_0$ estimates and one-third lower than can
  be achieved with extant CMB and BAO data.  We also analyze the
  impact of smaller uncertainty on measurements of $H_0$ as may be
  anticipated in the near term, in combination with CMB data from the
  Planck mission, and BAO data from the SDSS/BOSS program.  We
  demonstrate the possibility of a 5$\sigma$ detection for a fiducial
  neutrino mass of 0.1\,eV or a 95\,\% upper limit of 0.04\,eV for a
  fiducial of $m_\nu = 0\,\mbox{eV}$. These constraints are about
  50\,\% better than those achieved without external constraint. We
  further investigate the impact on modeling where the
    dark-energy equation of state is constant but not necessarily
    $-1$, or where a non-flat universe is allowed. In these cases,
  the next-generation accuracies of Planck, BOSS, and 1\,\%
  measurement of $H_0$ would all be required to obtain the limit
  $m_\nu < 0.05 - 0.06\,\mbox{eV}$ (95 \% C.L.)  for the fiducial of
  $m_\nu=0\,\mbox{eV}$.  The independence of systematics argues for
  pursuit of both BAO and $H_0$ measurements.

\end{abstract}

\thispagestyle{empty} 

\end{titlepage}

\setcounter{page}{1}

\section{Introduction} 
\label{sec:introduction}

Neutrino mass is now one of the most important targets in cosmology.
Since neutrino mass affects the evolution of the Universe in some
observable ways, a mass constraint can be obtained from the
cosmological data such as cosmic microwave background (CMB)
\cite{Ichikawa:2004zi, Spergel:2006hy, Fukugita:2006rm,
  Dunkley:2008ie, Komatsu:2008hk}, galaxy clustering
\cite{Tegmark:2006az}, Lyman-$\alpha$ forest \cite{Seljak:2006bg}, and
weak lensing data \cite{Tereno:2008mm,Ichiki:2008ye}.  Although
atmospheric, solar, reactor and accelerator neutrino oscillation
experiments can probe mass differences precisely ($|\Delta m^2_{21}|
= | m_2^2 - m_1^2|= 7.65^{+0.23}_{-0.20} \times 10^{-5}$~eV$^2$ and
$|\Delta m^2_{31}| = | m_3^2 - m_1^2 | = 2.40^{+0.12}_{-0.11} \times
10^{-3}$~eV$^2$, where $m_i$ is the mass of the $i$-th neutrino mass
eigenstate and errors are 1$\sigma$ \cite{Schwetz:2008er}), they are
not sensitive to absolute values.  Absolute mass may be inferred
from terrestrial experiments such as tritium beta decay and
neutrino-less double beta decay
\cite{Otten:2008zz,Elliott:2002xe,Vogel:2008sx}, but, cosmological
measurements from several observables have the potential to deliver
higher accuracy.

Analyses of the CMB data alone constrain neutrino masses, but there is
a relatively large degeneracy between neutrino masses and the Hubble
constant \cite{Ichikawa:2004zi}.  A combination of different data sets
can tighten limits.  CMB data and matter power spectrum measurements
can be usefully combined, since the suppression of the power spectrum
below the damping scale is dependent on the mass of free-streaming
neutrinos.  A neutrino mass constraint has previously been derived
\cite{Komatsu:2008hk} that combines CMB data with measurements of the
baryon acoustic oscillation (BAO) scale, supplemented by luminosity
distance measurements from type-Ia supernovae.  Since these non-CMB
observations directly characterize the distance scale, a high accuracy
prior on the Hubble constant can have a qualitatively similar effect,
as was first demonstrated for WMAP first-year data by
\cite{Ichikawa:2004zi}.

In this paper, we examine the extent to which priors on the Hubble
constant improve constraint on neutrino mass, without using matter
power spectrum data (see also \cite{Ichikawa:2007yb,Reid:2009nq}). 
Because systematic errors
for estimating the Hubble constant and the matter power spectrum are
different, these parallel routes to estimate neutrino mass are
important. We focus on the newly reported measurement of the Hubble
constant \cite{Riess:2009pu} that has a rather higher central value
and about two times smaller error than that of previous work
\cite{Freedman:2000cf}.  We also study the effect of tighter priors on
the Hubble constant, anticipating that the accuracy of estimates will
improve over time, at least through measurement of distance to a large
number of galaxies in the Hubble flow that host water maser sources
\cite{Macri:2006wm,Argon:2007ry,Greenhill:2009yi}. In this paper, we
denote the Hubble constant in unit of km/s/Mpc as $H_0$.

In the next section, we discuss the neutrino mass constraints using
the current CMB data set and the Hubble constant measurements.  We
demonstrate the degeneracy between neutrino masses and the Hubble
constant in analysis of CMB data alone, where we consider WMAP5,
ACBAR, CBI, BOOMERANG and QUAD.  We also demonstrate improvement in
the constraint through addition of the Hubble constant determination
of Ref.~\cite{Riess:2009pu}.  The degeneracy in the CMB mainly
originates from that in the angular diameter distance to last
scattering surface, so another observation probing such distances will
also be helpful to obtain the constraint. Hence we also make an
analysis that includes the BAO scale measurement released recently
\cite{Percival:2009xn}.  In Sec.~\ref{sec:planck}, future expected
limits for neutrino masses are presented, with particular attention to
the role of the prior on $H_0$.  For this purpose, we use projected
Planck CMB data and uncertainties in $H_0$ that may plausibly be
anticipated from estimation of distances to galaxies containing water
maser sources \cite{Macri:2006wm,Argon:2007ry,Greenhill:2009yi}.
Furthermore, we discuss the effect of future BAO data such as the
Baryon Oscillation Spectroscopic Survey (BOSS) on the neutrino mass
constraint.  In Sec.~\ref{sec:nu_mass_hubble} and \ref{sec:planck}, a
flat universe and a cosmological constant for dark energy are assumed.
However, in Sec.~\ref{sec:DE_curvature}, we investigate the constraint
for the case of a constant equation of state for dark energy and a
non-flat universe.  The final section summarizes the paper.

\section{Neutrino mass constraint in $\Lambda$CDM model}
\label{sec:nu_mass_hubble}

\subsection{Current CMB and $H_0$ measurements}

We first derive limits on neutrino mass using CMB data from WMAP and
other small angular scale measurements, including recently released
QUAD data \cite{Friedman:2009dt}. We demonstrate the degeneracy
between neutrino masses and the Hubble constant in the CMB data
analysis, which originates from geometry.  In light of this, we
examine how the mass limit is improved by adding a prior on $H_0$
obtained from the HST Key Project (KP) \cite{Freedman:2000cf} and a
newly available one from the Supernovae and $H_0$ for the Equation of
State (SHOES) program \cite{Riess:2009pu}.

\begin{table}[ht]
  \begin{center}
  \begin{tabular}{l||c}
  \hline
  \hline
  parameters & prior ranges\\
  \hline
  $\omega_b$ & $0.020 \to 0.025$ \\
  $\omega_c$ & $0.08 \to 0.14$ \\
  $\theta_s$ & $1.03 \to1.05$ \\
  $\tau$ &$0.01 \to 0.20$ \\
  $\Omega_k$ & ($-0.1 \to 0.1$) \\
  $\omega_\nu$ & $0 \to 0.02$ \\
  $w_X$ & ($-3 \to 0$) \\
  $n_s$ & $0.8 \to 1.1$ \\
  $\ln(10^{10}A_s)$ & $2.8\to3.5$ \\
  $A_\mathrm{SZ}$ & $0\to 4$ \\
  \hline
  $H_0$& $40\to100$ \\
  \hline
  \hline 
\end{tabular}
\caption{Adopted prior ranges for cosmological parameters described in
  the main text.  In addition to top-hat priors on the primary
  parameters, we also impose an additional prior $H_0\in [40,100]$,
  which is hardcoded in {\tt CosmoMC} by default.  Also note that the
  prior ranges for $\Omega_k$ and $w_X$, shown in the parentheses, are
  adopted only when they are varied in Section
  \ref{sec:DE_curvature}.}
  \label{table:priors}
\end{center}
\end{table}
\begin{table}
  \begin{center}
  \begin{tabular}{lrrr}
    \hline
    \hline
    parameters & CMB & CMB+KP & CMB+SHOES  \\
    \hline
    $\omega_b\times10^2$ & 
    $2.258^{+0.053}_{-0.066}$ & 
    $2.267^{+0.057}_{-0.055}$ & 
    $2.291^{+0.055}_{-0.048}$ \\
    $\omega_c$ & 
    $0.1095^{+0.0050}_{-0.0056}$ & 
    $0.1087^{+0.0043}_{-0.0055}$ & 
    $0.1064^{+0.0048}_{-0.0039}$ \\
    $\theta_s\times10^2$ & 
    $104.14^{+0.21}_{-0.23}$ & 
    $104.16^{+0.20}_{-0.24}$ & 
    $104.21^{+0.23}_{-0.19}$ \\
    $\tau$ & 
    $0.087^{+0.015}_{-0.017}$ & 
    $0.088^{+0.015}_{-0.017}$ & 
    $0.090^{+0.015}_{-0.018}$ \\
    $\omega_\nu$ & 
    $<0.013 ~(95\%)$ & 
    $<0.011 ~(95\%)$ & 
    $<0.0065 ~(95\%)$ \\
    $n_s$ & 
    $0.952^{+0.018}_{-0.014}$ & 
    $0.956^{+0.015}_{-0.013}$ & 
    $0.963^{+0.012}_{-0.012}$ \\
    $\ln[10^{10}A_s]$ & 
    $3.049^{+0.039}_{-0.035}$ & 
    $3.050^{+0.038}_{-0.036}$ & 
    $3.050^{+0.041}_{-0.033}$ \\
    \hline
    age~[Gyr] & 
    $13.94^{+0.22}_{-0.28}$ & 
    $13.88^{+0.15}_{-0.27}$ & 
    $13.73^{+0.10}_{-0.17}$ \\
    $\Omega_m$ & 
    $0.308^{+0.040}_{-0.068}$ & 
    $0.292^{+0.033}_{-0.054}$ & 
    $0.258^{+0.021}_{-0.029}$ \\
    $z_\mathrm{re}$ & 
    $10.5^{+1.3}_{-1.3}$ & 
    $10.5^{+1.3}_{-1.3}$ & 
    $10.6^{+1.3}_{-1.3}$ \\
    $H_0$ & 
    $67.4^{+4.9}_{-4.4}$ & 
    $68.6^{+4.8}_{-3.0}$ & 
    $71.6^{+2.8}_{-2.1}$ \\
    $m_\nu$~[eV] & 
    $<0.41 ~(95\%)$ & 
    $<0.34 ~(95\%)$ & 
    $<0.20 ~(95\%)$ \\
    \hline
    \hline
  \end{tabular}
  \caption{Constraints on the $\Lambda$CDM+$m_\nu$ model from current
    CMB data (WMAP5+ACBAR+CBI+BOOMERANG+QUAD) with and without priors
    on $H_0$.  We adopt two different priors on $H_0$ from direct
    measurements, the KP prior of $H_0=72\pm 8$ \cite{Freedman:2000cf}
    and the SHOES prior of $H_0=74.2\pm3.6$ \cite{Riess:2009pu}.  In
    the table, errors are at $68$~\% C.L., except for upper limits at
    $95$~\% C.L. on $\omega_\nu$ and $m_\nu$, which are not bounded
    from below.  }
  \label{table:current_LCDM1}
  \end{center}
\end{table}

\begin{table}
  \begin{center}
  \begin{tabular}{lrrr}
    \hline
    \hline
    parameters & CMB+BAO & CMB+BAO+KP & CMB+BAO+SHOES  \\
    \hline
    $\omega_b\times10^2$ & $2.264^{+0.047}_{-0.057}$ & $2.266^{+0.053}_{-0.050}$ & $2.279^{+0.047}_{-0.053}$ \\
    $\omega_c$ & $0.1093^{+0.0038}_{-0.0041}$ & $0.1092^{+0.0038}_{-0.0041}$ & $0.1090^{+0.0038}_{-0.0037}$ \\
    $\theta_s\times10^2$ & $104.13^{+0.22}_{-0.21}$ & $104.13^{+0.22}_{-0.21}$ & $104.16^{+0.24}_{-0.17}$ \\
    $\tau$ & $0.087^{+0.016}_{-0.015}$ & $0.088^{+0.016}_{-0.015}$ & $0.088^{+0.015}_{-0.016}$ \\
    $\omega_\nu$ & $<0.0090 ~(95\%)$ & $<0.0086 ~(95\%)$ & $<0.0071 ~(95\%)$ \\
    $n_s$ & $0.955^{+0.011}_{-0.012}$ & $0.955^{+0.011}_{-0.012}$ & $0.958^{+0.011}_{-0.011}$ \\
    $\ln[10^{10}A_s]$ & $3.050^{+0.035}_{-0.037}$ & $3.051^{+0.035}_{-0.038}$ & $3.054^{+0.040}_{-0.033}$ \\
    \hline
    age~[Gyr] & $13.88^{+0.14}_{-0.16}$ & $13.87^{+0.14}_{-0.16}$ & $13.80^{+0.10}_{-0.15}$ \\
    $\Omega_m$ & $0.291^{+0.018}_{-0.025}$ & $0.289^{+0.019}_{-0.023}$ & $0.277^{+0.014}_{-0.020}$ \\
    $z_\mathrm{re}$ & $10.4^{+1.2}_{-1.4}$ & $10.4^{+1.2}_{-1.4}$ & $10.5^{+1.2}_{-1.4}$ \\
    $H_0$ & $68.4^{+2.3}_{-1.6}$ & $68.6^{+2.1}_{-1.7}$ & $69.8^{+1.7}_{-1.6}$ \\
    $m_\nu$~[eV] & $<0.28 ~(95\%)$ & $<0.27 ~(95\%)$ & $<0.22 ~(95\%)$ \\
    \hline
    \hline
  \end{tabular}
  \caption{Constraints on the $\Lambda$CDM+$m_\nu$ model from current
    CMB and the BAO data, with and without priors on $H_0$.  }
  \label{table:current_LCDM2}
  \end{center}
\end{table}

\begin{figure}[t]
\begin{center}
  \scalebox{1.5}{\includegraphics{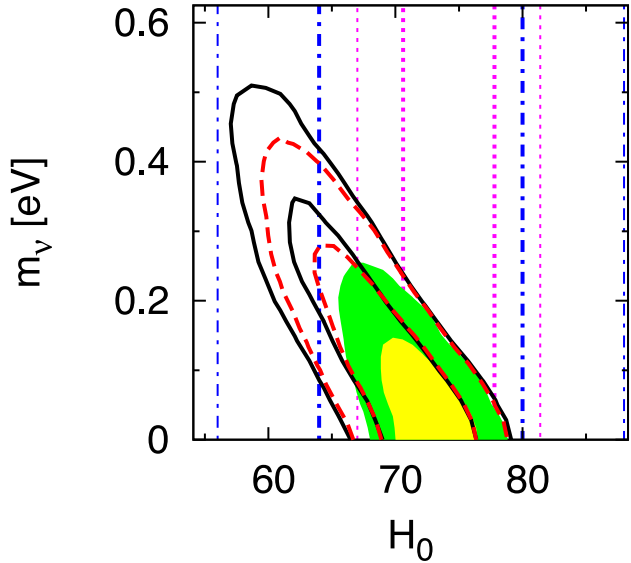}}
  \caption{Regions of 68\% and 95\% C.L. from current CMB data with no
    prior on $H_0$ (black solid line), with the KP prior $H_0 = 72
    \pm8$ \cite{Freedman:2000cf} (red dashed line) and with the SHOES
    prior $H_0 = 74.2 \pm3.6$ \cite{Riess:2009pu} (green and yellow
    shaded regions).  Vertical lines show priors on $H_0$ at 1$\sigma$
    and 2$\sigma$ level from the KP (blue dot-dashed line) and the
    SHOES (magenta dotted line) measurements. Inner (outer) contours
    and shaded regions show regions of 68\% (95\%) C.L.  }
\label{fig:mnu_h_current}
\end{center}
\end{figure}

We begin by showing a neutrino mass constraint from current CMB data
sets alone.  In the analysis of this section, we assume a flat
universe and a cosmological constant for dark energy.  For neutrinos,
we assume that there are three generations of active neutrinos with
equal mass $m_\nu$, so that $\sum m_\nu=3m_\nu$.  We also assume there
is no lepton asymmetry in the Universe\footnote{
  The constraint on neutrino masses including lepton asymmetry has
  been investigated in \cite{Popa:2008tb,Shiraishi:2009fu}.
}.  We used CMB data from WMAP5 \cite{Nolta:2008ih,Dunkley:2008ie},
ACBAR\cite{Kuo:2006ya}, CBI \cite{Sievers:2005gj},
BOOMERANG\cite{Jones:2005yb,Piacentini:2005yq,Montroy:2005yx} and QUAD
\cite{Friedman:2009dt}.  We performed a Monte Carlo analysis with some
modifications to {\tt MultiNest} \cite{Feroz:2008xx}, integrated in
{\tt CosmoMC} \cite{Lewis:2002ah}. In the analysis, we explored 7+1
dimensional parameter space ($\omega_b, \omega_c, \theta_s, \tau,
\omega_\nu, n_s, A_s, A_{\rm SZ}$). Their prior ranges are shown in
Table~\ref{table:priors}. Here, $\omega_b = \Omega_b h^2$, $\omega_c
=\Omega_c h^2$ and $\omega_\nu =\Omega_\nu h^2$ are energy density
parameters for baryon, cold dark matter and massive neutrinos,
respectively, where $\Omega$ denotes energy density normalized by the
critical energy density and $h$ is defined by $H_0=100h$.  As well,
$\theta_s$ is the acoustic peak scale, $\tau$ is the optical depth of
reionization, $n_s$ is the spectral index of primordial power
spectrum, $A_s$ is the amplitude of primordial fluctuations, and
$A_{\rm SZ}$ is the amplitude of thermal SZ effect, which is
normalized to the $C_l^{SZ}$ template from Ref.~\cite{Komatsu:2002wc}.
The neutrino mass is related to the density parameter by $\omega_\nu =
3 m_\nu/(93.8$\,eV). The cosmological parameters estimated from this
CMB data set are summarized in the left column of
Table~\ref{table:current_LCDM1}, and the 95\% confidence level (C.L.)
upper limit on the neutrino mass is found to be $m_\nu <
0.41$\,eV. This is almost same as the constraint from WMAP5 data alone
($m_\nu < 0.43$\,eV) \cite{Komatsu:2008hk}.

Before we closely look at how the prior on $H_0$ affects the
constraint on $m_\nu$, we discuss the degeneracy between $m_\nu$ and
$H_0$.  There is a strong negative correlation between $m_\nu$ and
$H_0$ (Fig.\,\ref{fig:mnu_h_current}), where correlation coefficients
between $m_\nu$ and $(\omega_b, \omega_c, \tau, n_s, H_0)$ are
$(-0.43, 0.22, -0.13, -0.59, -0.86)$.  The negative correlation may be
understood in terms of the position of the acoustic peaks of the CMB
power spectrum, which is given by the ratio of sound horizon\footnote{
  The sound horizon $r_s$ at the recombination epoch $a_{\rm rec}$
  is given by
\[
r_s (a_{\rm rec}) = \int_0^{a_{\rm rec}} \frac{c_s}{a^2H} da
\]
where $c_s$ is the sound speed of the photon-baryon fluid.
} to the distance to the last scattering surface, $d_{\rm lss}$. When
$m_\nu$ increases for given $\omega_c$, there is a larger
non-relativistic component in the present universe (which makes
expansion faster), and hence $d_{\rm lss}$ is smaller. Meanwhile,
$d_{\rm lss}$ is inversely proportional to $H_0$, and thus larger
$H_0$ gives smaller $d_{\rm lss}$. Therefore, the effect of increasing
$m_\nu$ is cancelled by decreasing $H_0$. Note that the sound horizon
could be affected by $m_\nu$ if it is so large that neutrinos become
non-relativistic before the epoch of recombination\footnote{
  The epoch when neutrinos become non-relativistic can be roughly
  evaluated by taking the time of $\langle p_\nu \rangle \sim m_\nu$
  with $\langle p_\nu \rangle$ being the average momentum of neutrinos
  which is roughly $\sim 3 T_\nu$ where $T_\nu$ is the temperature of
  neutrinos.  This gives
\begin{equation*}
1 + z_{\nu {\rm NR}} \simeq 2 \times 10^3 \left( \frac{m_{\nu}}{1\,\mbox{eV}} \right).
\end{equation*}
}.  However, such large masses would modify the heights of the CMB
acoustic peaks in a characteristic manner, which is excluded by
current CMB data. (This fact gives the upper bound on $m_\nu$
described in Table\,\ref{table:current_LCDM1}.)

This argument can be quantified by calculating how the first peak
position of CMB power spectrum $l_1$ responds as $m_\nu$ or $H_0$
vary. (For a detailed description of how this can be done, see
\cite{Ichikawa:2004zi,Ichikawa:2007js,Ichikawa:2008pz}.)  Note that
$l_1$ becomes smaller (moves toward larger scales) when $d_{\rm lss}$
decreases.  We find $\Delta l_1/\Delta m_\nu = -19.7$ and $\Delta
l_1/\Delta H_0 = -0.643$. Therefore, the direction with no change in
$l_1$ is $\Delta m_\nu/\Delta H_0 = 0.033$, which roughly describes
the slope of the correlation in Fig.~\ref{fig:mnu_h_current}.

A tighter upper limit on $m_\nu$ flows directly from a lower bound on
$H_0$, independent of CMB data \cite{Ichikawa:2004zi}.  The results
for two priors, from the KP $H_0=72 \pm 8$\,\cite{Freedman:2000cf} and
SHOES $H_0=74.2 \pm 3.6$\,\cite{Riess:2009pu} are presented in
Table~\ref{table:current_LCDM1}, where the upper bound on neutrino
mass is as low as $m_\nu < 0.20$\,eV\footnote{ The latter case is
also investigated in Ref.~\cite{Reid:2009nq} and our result is
consistent with theirs.  }.

\subsection{Current CMB and BAO measurements}

Measurement of the BAO scale also fixes geometry and may be used to
constraint $m_\nu$.  We follow up with an analysis of $m_\nu$ in the
context of current CMB data, augmented by the recent BAO data from
\cite{Percival:2009xn}.  In Table~\ref{table:current_LCDM2},
constraints on the cosmological parameters are shown for the analyses
of CMB+BAO, CMB+BAO+KP and CMB+BAO+SHOES.  As anticipated, the
constraints on $m_\nu$ from the CMB+BAO analyses are improved as well.
However, some cautionary words are in order.  The results indicate a
tighter upper limit for CMB+SHOES analysis than CMB+BAO analysis.
This comes from the fact that BAO data prefers a relatively low
value of $H_0$, which is outside the SHOES confidence interval.
The joint CMB+BAO+SHOES analysis provides a less tighter limit on
$m_\nu$ compared to that of CMB+SHOES, and it does not resolve
friction between the contributing data and should be interpreted
with care.  In light of their different sources of systematic
error, it may be more advisable, for the time being, to interpret the
SHOES and BAO data separately.

So far, we have investigated the cosmological constraints on $m_\nu$
using currently available measurements of CMB, BAO and $H_0$.
However, measurements of these are expected to be more precise with
new observations in the near future such as Planck, BOSS and the $H_0$
measurement mentioned in the introduction. Thus, in the next section,
we consider future constraints on the neutrino mass based on CMB, BAO
and $H_0$.

\section{Impact of improved  cosmological data on $\Lambda$CDM estimation of $m_\nu$}
\label{sec:planck}

\begin{table}[t]
  \begin{center}
  \begin{tabular}{l||c|c}
  \hline
  \hline
  parameters & fiducial values & prior ranges\\
  \hline
  $\omega_b$ & $0.02273$ & $0.20\to0.25$\\
  $\omega_c$ & $0.1099[0.1067]$ & $0.08\to0.14$ \\
  $\theta_s$ & $1.04062[1.04684]$ & $1.03\to1.05$\\
  $\tau$ & $0.87$ & $0.01\to 0.2$ \\
  $\Omega_k$ & $0$ & ($-0.1\to0.1$) \\
  $\omega_\nu$ & $0[0.0032]$ & $0\to0.02$\\
  $w_X$ & $-1$ & ($-2\to0$) \\
  $n_s$ & $0.963$ & $0.8\to1.1$ \\
  $\ln(10^{10}A_s)$ & $3.063$ & $2.8\to3.5$ \\
  \hline
  $H_0$ & $71.9$ & --- \\
  $m_\nu$~[eV] & 0[0.1] & --- \\
  \hline
  \hline 
\end{tabular}
\caption{Fiducial cosmological parameters and prior ranges used for
  predictive analysis. The fiducial values here correspond to the mean
  values of the flat $\Lambda$CDM model from analysis of the WMAP5
  data alone.  Regarding the fiducial neutrino mass, we consider two
  cases, $m_\nu=0$ and $0.1$~eV, assuming degenerate mass hierarchy
  (three neutrino species have the same mass). We fix the energy
  density of dark matter and the Hubble constant to be $\omega_{\rm
    dm}=0.1099$ and $H_0=71.9$, and hence $\omega_c$ and $\theta_s$
  vary with $m_\nu$.  Values of $\omega_\nu$, $\omega_c$ and
  $\theta_s$ are shown for $m_\nu=0$ [0.1] eV.  Also note that the
  prior ranges for $w_X$ and $\Omega_k$, shown with parentheses, are
  adopted only when they are varied in Section
  \ref{sec:DE_curvature}.}
  \label{table:fiducial}
\end{center}
\end{table}

\begin{table}
  \begin{center}
  \begin{tabular}{lrr}
    \hline
    \hline
    redshifts & $\sigma_{d_A(z)} (\%)$ & $\sigma_{H(z)} (\%)$ \\
    \hline
    $0.35$ & $1.0$ & $1.8$\\
    $0.6$ & $1.1$ & $1.7$ \\
    $2.5$ & $1.5$ & $1.5$ \\
    \hline
    \hline
  \end{tabular}
  \caption{Sensitivities of the BOSS survey~\cite{Schlegel:2009hj}
    adopted in the analysis ({\tt
      http://www.sdss3.org/collaboration/description.pdf}).  Expected
    error~(\%) for angular diameter distance $d_A(z)$ and the Hubble
    expansion rate $H(z)$ are presented. Correlations among
      errors are assumed to be absent. }
  \label{table:BOSS}
  \end{center}
\end{table}

\begin{figure}[h]
\begin{center}
   \hspace{-10mm}
  \resizebox{80mm}{!}{\includegraphics{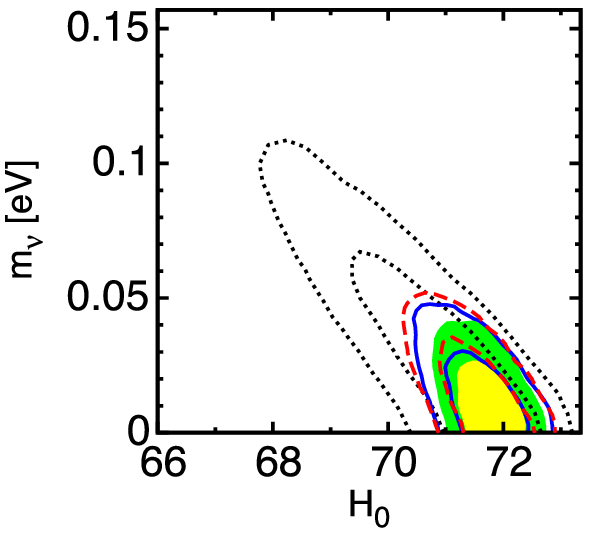}}
     \hspace{5mm}
  \resizebox{80mm}{!}{\includegraphics{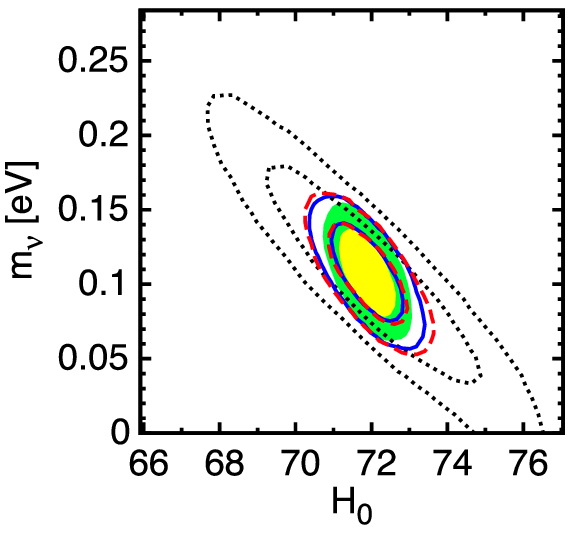}}
  \caption{Regions of 68\% and 95\%\,C.L. on the $m_\nu$--$H_0$ plane
    from projected data of Planck with no other data (black dotted
    line), with $H_0$ at 1\% accuracy (red dashed line), with BAO
    scales from BOSS (blue solid line) and both $H_0$ at 1\% and BAO
    scales from BOSS combined (green and yellow shaded regions).
    Inner (outer) contours and shaded regions show 68\% (95\%)
    C.L. limits. The fiducial values for $m_\nu$ are taken to be
    $m_\nu =0$\,eV (left panel) and $0.1$\,eV (right panel). }
\label{fig:mnu_h_future}
\end{center}
\end{figure}
\begin{table}
  \begin{center}
  \begin{tabular}{lrrrrr}
    \hline
    \hline
    parameters & Planck & Planck+$H_0$ & Planck+BOSS & Planck+BOSS+$H_0$ \\
    \hline
    $\omega_b\times10^2$ & $2.267^{+0.013}_{-0.013}$ & $2.273^{+0.011}_{-0.011}$ & $2.274^{+0.012}_{-0.010}$ & $2.275^{+0.012}_{-0.010}$ \\
    $\omega_c$ & $0.1101^{+0.0009}_{-0.0012}$ & $0.10949^{+0.00080}_{-0.00080}$ & $0.10946^{+0.00078}_{-0.00085}$ & $0.10936^{+0.00078}_{-0.00072}$ \\
    $\theta_s\times10^2$ & $104.055^{+0.026}_{-0.027}$ & $104.067^{+0.023}_{-0.024}$ & $104.067^{+0.023}_{-0.023}$ & $104.069^{+0.023}_{-0.023}$ \\
    $\tau$ & $0.0878^{+0.0040}_{-0.0047}$ & $0.0883^{+0.0038}_{-0.0047}$ & $0.0882^{+0.0039}_{-0.0048}$ & $0.0883^{+0.0038}_{-0.0049}$ \\
    $\omega_\nu$ & $<0.0028 ~(95\%)$ & $<0.0014 ~(95\%)$ & $<0.0012 ~(95\%)$ & $<0.0011 ~(95\%)$ \\
    $n_s$ & $0.9617^{+0.0036}_{-0.0028}$ & $0.9632^{+0.0024}_{-0.0031}$ & $0.9632^{+0.0029}_{-0.0026}$ & $0.9635^{+0.0026}_{-0.0027}$ \\
    $\ln[10^{10}A_s]$ & $3.0647^{+0.0070}_{-0.0096}$ & $3.0643^{+0.0072}_{-0.0090}$ & $3.0640^{+0.0076}_{-0.0093}$ & $3.0640^{+0.0075}_{-0.0093}$ \\
    \hline
    age[Gyr] & $13.757^{+0.035}_{-0.068}$ & $13.713^{+0.024}_{-0.026}$ & $13.710^{+0.019}_{-0.025}$ & $13.706^{+0.018}_{-0.020}$ \\
    $\Omega_m$ & $0.2688^{+0.0073}_{-0.0142}$ & $0.2597^{+0.0046}_{-0.0056}$ & $0.2591^{+0.0048}_{-0.0047}$ & $0.2580^{+0.0037}_{-0.0042}$ \\
    $z_\mathrm{re}$ & $10.53^{+0.31}_{-0.42}$ & $10.52^{+0.33}_{-0.37}$ & $10.51^{+0.32}_{-0.39}$ & $10.51^{+0.33}_{-0.38}$ \\
    $H_0$ & $70.64^{+1.24}_{-0.85}$ & $71.51^{+0.54}_{-0.46}$ & $71.58^{+0.48}_{-0.44}$ & $71.68^{+0.35}_{-0.41}$ \\
    $m_\nu$~[eV] & $<0.088~(95\%)$ & $<0.042~(95\%)$ & $<0.038~(95\%)$ & $<0.034~(95\%)$ \\
    \hline
    \hline
  \end{tabular}
  \caption{Constraints on the $\Lambda$CDM+$m_\nu$ model from analysis
    of Planck, Planck+$H_0$, Planck+BOSS and Planck+BOSS+$H_0$ data.
    The fiducial value for the neutrino mass is taken to be
    $m_\nu=0$~eV.  We assume a prior on $H_0$ with 1\% uncertainty
    (1$\sigma$) provided by a future direct measurement.  }
  \label{table:Planck0}
  \end{center}
\end{table}

\begin{table}
  \begin{center}
  \begin{tabular}{lrrrrr}
    \hline
    \hline
    parameters & Planck & Planck+$H_0$ & Planck+BOSS & Planck+BOSS+$H_0$ \\
    \hline
    $\omega_b\times10^2$ & $2.275^{+0.014}_{-0.014}$ & $2.275^{+0.012}_{-0.011}$ & $2.274^{+0.012}_{-0.011}$ & $2.274^{+0.012}_{-0.011}$ \\
    $\omega_c$ & $0.1071^{+0.0011}_{-0.0012}$ & $0.10713^{+0.00082}_{-0.00085}$ & $0.10713^{+0.00077}_{-0.00092}$ & $0.10713^{+0.00077}_{-0.00086}$ \\
    $\theta_s\times10^2$ & $104.677^{+0.031}_{-0.027}$ & $104.676^{+0.022}_{-0.026}$ & $104.675^{+0.026}_{-0.023}$ & $104.675^{+0.025}_{-0.022}$ \\
    $\tau$ & $0.0877^{+0.0044}_{-0.0046}$ & $0.0877^{+0.0043}_{-0.0045}$ & $0.0879^{+0.0044}_{-0.0047}$ & $0.0879^{+0.0045}_{-0.0046}$ \\
    $\omega_\nu$ & $<0.0063~(95\%)$ & $0.00345^{+0.00069}_{-0.00064}$ & $0.00348^{+0.00065}_{-0.00062}$ & $0.00347^{+0.00054}_{-0.00056}$ \\
    $n_s$ & $0.9642^{+0.0038}_{-0.0038}$ & $0.9642^{+0.0027}_{-0.0031}$ & $0.9642^{+0.0027}_{-0.0029}$ & $0.9643^{+0.0028}_{-0.0027}$ \\
    $\ln[10^{10}A_s]$ & $3.0653^{+0.0084}_{-0.0088}$ & $3.0654^{+0.0082}_{-0.0089}$ & $3.0657^{+0.0076}_{-0.0099}$ & $3.0657^{+0.0075}_{-0.0100}$ \\
    \hline
    age[Gyr] & $13.664^{+0.096}_{-0.092}$ & $13.664^{+0.033}_{-0.037}$ & $13.667^{+0.035}_{-0.030}$ & $13.666^{+0.028}_{-0.024}$ \\
    $\Omega_m$ & $0.258^{+0.017}_{-0.017}$ & $0.2579^{+0.0066}_{-0.0062}$ & $0.2581^{+0.0058}_{-0.0058}$ & $0.2580^{+0.0041}_{-0.0049}$ \\
    $z_\mathrm{re}$ & $10.48^{+0.35}_{-0.40}$ & $10.48^{+0.34}_{-0.40}$ & $10.49^{+0.38}_{-0.38}$ & $10.49^{+0.38}_{-0.37}$ \\
    $H_0$ & $71.9^{+1.7}_{-1.8}$ & $71.92^{+0.60}_{-0.71}$ & $71.88^{+0.55}_{-0.64}$ & $71.89^{+0.46}_{-0.46}$ \\
    $m_\nu$~[eV]  & $0.107^{+0.045}_{-0.049}$ & $0.107^{+0.021}_{-0.020}$ & $0.108^{+0.020}_{-0.019}$ & $0.107^{+0.016}_{-0.017}$ \\
    \hline
    \hline
  \end{tabular}
  \caption{Same table as in Table~\ref{table:Planck0}, but for the
    fiducial neutrino mass $m_\nu=0.1$~eV.  }
  \label{table:Planck1}
  \end{center}
\end{table}

Next-generation cosmological data will tighten the limits that can be
placed on $m_\nu$.  We investigate the impact of data from the Planck
CMB mission, a high-accuracy prior on $H_0$, as may be achieved from
study of water masers
\cite{Macri:2006wm,Argon:2007ry,Greenhill:2009yi}\footnote{
    $H_0$ may also be constrained tightly by the gravitational lens time delays
    \cite{Oguri:2006qp,Coe:2009wt,Paraficz:2009xj}. }, and analysis of the
SDSS BAO survey \cite{Schlegel:2009hj}.

We again perform a Monte Carlo analysis to obtain a constraint on
$m_\nu$.  Regarding the expected data from Planck, we analyze it for
multipoles $l<2500$, obviating the need to consider the SZ
effect. Thus, we do not include the parameter $A_{\rm SZ}$ in the
analysis here.  We make use of the lensing of CMB in the likelihood
calculation, which is effective at constraining masses of neutrino in
the future observations of CMB~\cite{Perotto:2006rj}.  For the
analyses in this section, we assume fiducial cosmological parameters
summarized in Table~\ref{table:fiducial}.  For the projected Planck
data, we adopt the instrumental specifications found in
Ref.~\cite{Perotto:2006rj}.  Regarding a future BAO data set, we adopt
predicted performance of the BOSS survey, which will measure the
angular diameter distance $d_A(z)$ and the Hubble expansion rate
$H(z)$ of the Universe over a broad range of redshifts
(Table~\ref{table:BOSS}).  Roughly speaking, the BOSS survey adopted
here measures geometry of the Universe both in transverse and the line
of sight direction with uncertainties of about 1 \% level at redshifts
$z=0.35$, $0.6$ and $2.5$.  For the mass of neutrino, we assume two
fiducial values of $m_\nu =0~\mbox{eV}$ and $0.1~\mbox{eV}$.  The
fiducial value of 0.1\,eV is motivated from the KATRIN neutrino mass
direct search experiment, which will achieve a sensitivity of $m_\nu <
0.2$\,eV (95\%\,C.L.) within the next decade \cite{Steidl:2009hx}.

In Tables~\ref{table:Planck0} and \ref{table:Planck1}, we summarize
the projected constraints on cosmological parameters for the cases
with the fiducial values of $m_\nu=0$ and $0.1\,\mbox{eV}$ with and
without 1\, \% prior on $H_0$.  The cases where we include the BAO
data or a combination of Planck, BAO and $H_0$are also shown.  In
Fig.~\ref{fig:mnu_h_future}, contours of 68\,\% and 95\,\%\,C.L.
constraints are depicted for two different fiducial values for the
neutrino mass $m_\nu =0$ and $0.1~\mbox{eV}$. In both cases, the
degeneracy between $m_\nu$ and $H_0$ is significant for analyses of
CMB data alone.  However, imposing a Gaussian prior for the Hubble
constant with the error of 1\,\% improves the constraint
significantly. For the case of the fiducial value of
$m_\nu=0~\mbox{eV}$, 95\%\,C.L. limit for Planck data is $m_\nu <
0.088~\mbox{eV}$, which improves to $m_\nu < 0.042~\mbox{eV}$ for a
prior on $H_0$ with 1\,\% error.  When we assume $m_\nu =
0.1~\mbox{eV}$ as a fiducial value, the neutrino mass is constrained
to be $m_\nu < 0.197~\mbox{eV}$ if no prior on $H_0$ is adopted. Very
interestingly, when we impose the prior on the Hubble constant with
1\,\% error, the constraint becomes $m_\nu =
0.107^{+0.021}_{-0.020}~\mbox{eV}$ (68\%\,C.L.), which indicates that
we can detect neutrino masses at about $5 \sigma$ level if $m_\nu =
0.1~\mbox{eV}$.  Here it should be noted that even if we vary the
fiducial value of $H_0$ by 10\,\%, the limit is almost unchanged.  As
seen from Fig.~\ref{fig:mnu_h_future}, when we assume no prior on
$H_0$, the massless neutrino is still allowed at 2$\sigma$ level for
this case.  This analysis shows that an accurate determination of
$H_0$ may lead us towards the detection of the absolute mass of
neutrino from cosmology and therefore this direction research should
be pursued along with other methods.

The degree of constraint on $m_\nu$ is in fact a continuous function
of measurement uncertainty on $H_0$, and thus far 1\,\% has been
illustrative.  For a fiducial mass $m_\nu=0\,\mbox{eV}$ and an
expected error $\Delta H_0/H_0 ~(1\sigma)$, the constraints on $m_\nu$
can be fitted as, by marginalizing over other cosmological
parameters\begin{equation} m_\nu<0.045\left(\frac{\Delta
      H_0}{H_0}[\%]\right)^{0.36}~\mbox{(95\,\%\,C.L.)}. \label{eq:fit0}
\end{equation}
On the other hand, for a fiducial $m_\nu=0.1$~eV, the constraints can
be fitted as
\begin{equation}
m_\nu=0.1\pm0.022\left(\frac{\Delta H_0}{H_0}[\%]\right)^{0.41}~\mbox{(68\,\%\,C.L.)}. \label{eq:fit01}
\end{equation}
Eqs.~(\ref{eq:fit0}) and (\ref{eq:fit01}) are valid for the range
$1\%\le\Delta H_0/H_0\le5\%$ in the $\Lambda$CDM+$m_\nu$ model.

When the BAO data from BOSS are combined with Planck data, similar
limits can be obtained as seen from Tables~\ref{table:Planck0} and
\ref{table:Planck1}.  These analyses indicate that observations of the
geometric distances can very usefully constrain the neutrino mass.  As
demonstrated, in the near term, accurate observations of $H_0$ and the
BAO scale would give almost comparable limits when these are combined
with CMB data. Needless to say, when both an $H_0$ measurement and BAO
data are combined with CMB, we can obtain a tighter constraint on
$m_\nu$ than that with CMB data alone.

As a closing remark of this section, we note that 5$\sigma$ detection
of the neutrino mass of $m_\nu =0.1\,\mbox{eV}$ requires a 1\,\%
accuracy of $H_0$ measurement.  When the accuracy is 3\,\%, it becomes
3$\sigma$.  Thus for the detection of the neutrino mass with high
significance, a very accurate measurement such as that with 1\,\%
error would be necessary.

\section{Effects of dark energy equation of state and the curvature of the Universe}
\label{sec:DE_curvature}

So far, we have investigated constraints on neutrino masses, assuming
a flat $\Lambda$CDM cosmological model.  We have discussed that there
exists a degeneracy between $m_\nu$ and $H_0$ in CMB power spectrum,
but by adopting the prior on $H_0$ and/or combining the BAO data, we
can break the degeneracy to some extent and obtain a tighter
constraint on $m_\nu$.  However, we note that the extent of the
degeneracy depends on distance scale, and this is affected by the
equation of state for dark energy and the curvature of the Universe, 
which we have previously fixed to be $w_X=-1$ and
$\Omega_k=0$.  Therefore here we relax these assumptions and
investigate how the variations of $w_X$ and $\Omega_k$ affect the
determination of $m_\nu$, paying particular attention to the
combination of present-day CMB data and constraints on $H_0$ and the
BAO scale.  Furthermore, as we did in the previous section, we also
investigate an attainable bound on $m_\nu$ in the near future using
the projected characteristics of Planck data and anticipated
constraints on $H_0$ and the BAO scale.

In Tables~\ref{table:current_wCDM}--\ref{table:current_owCDM}, we
summarized the constraints on cosmological parameters for the cases
where (i)~the equation of state for dark energy $w_X\neq -1$ is
allowed (but constant in time) and a flat universe is assumed
($w$CDM model), (ii)~$w_X=-1$ but a non-flat universe is allowed
($\Omega\Lambda$CDM model) and (iii)~$w_X\neq -1$ and a
non-flat universe are allowed ($\Omega w$CDM model), respectively.
Since the geometric degeneracy between $\Omega_k$ and $H_0$ is
particularly very large in the CMB analysis, we do not obtain
meaningful cosmological constraints on them through analysis of CMB
data alone for non-flat cases, such as $\Omega\Lambda$CDM and $\Omega
w$CDM models.  (Hence these CMB analyses are not presented in
Tables~\ref{table:current_oLCDM}--\ref{table:current_owCDM}.)
Although this problem is mitigated somewhat through combinations of an
$H_0$ prior and BAO data with CMB data, a degeneracy remains among
$H_0$, $w_X$, and $\Omega_k$ because each affects the distance scales.
Thus even if we include the geometry measurements such as $H_0$ and
the BAO scale, the constraint on $m_\nu$ is not so improved when $w_X$
and/or $\Omega_k$ are varied since the change of the geometric
distance made by varying $m_\nu$ by the amount
$\mathcal{O}(0.1)\,\mbox{eV}$ can be easily compensated by a
relatively small adjustment of $w_X$ and $H_0$, and then the neutrino
mass effectively decouples from the degeneracy in the geometric
distance (Tables~\ref{table:current_wCDM}--\ref{table:current_owCDM}).
This is in contrast to analysis of the flat $\Lambda$CDM model, where
the degeneracy involves just two quantities rather than four.
Nonetheless, the combination of distance scale measurements
substantively improve limits on $w_X$ and $\Omega_k$.  Furthermore, we
note that for $w$CDM model, the limit on $m_\nu$ from CMB+BAO data is
looser than that for CMB alone. This is because the BAO data
\cite{Percival:2009xn} favor a smaller value for $w_X$, and $w_X$ and
$m_\nu$ are negatively correlated.

\begin{table}
  \begin{center}
  \begin{tabular}{lrrrr}
    \hline
    \hline
    parameters & CMB & CMB+SHOES & CMB+BAO & CMB+BAO+SHOES  \\
    \hline
    $\omega_b\times10^2$ & $2.248^{+0.054}_{-0.066}$ & $2.249^{+0.058}_{-0.060}$ & $2.231^{+0.049}_{-0.064}$ & $2.230^{+0.049}_{-0.060}$ \\
    $\omega_c$ & $0.1103^{+0.0048}_{-0.0058}$ & $0.1103^{+0.0052}_{-0.0051}$ & $0.1121^{+0.0051}_{-0.0051}$ & $0.1123^{+0.0050}_{-0.0043}$ \\
    $\theta_s\times10^2$ & $104.12^{+0.23}_{-0.20}$ & $104.13^{+0.25}_{-0.19}$ & $104.10^{+0.21}_{-0.22}$ & $104.10^{+0.22}_{-0.22}$ \\
    $\tau$ & $0.086^{+0.015}_{-0.017}$ & $0.086^{+0.013}_{-0.018}$ & $0.084^{+0.015}_{-0.016}$ & $0.084^{+0.015}_{-0.015}$ \\
    $\omega_\nu$ & $<0.014 ~(95\%)$ & $<0.013 ~(95\%)$ & $<0.015 ~(95\%)$ & $<0.015 ~(95\%)$ \\
    $w_X$ & $-1.29^{+0.55}_{-0.47}$ & $-1.27^{+0.27}_{-0.15}$ & $-1.41^{+0.52}_{-0.24}$ & $-1.38^{+0.32}_{-0.19}$\\
    $n_s$ & $0.949^{+0.016}_{-0.016}$ & $0.948^{+0.018}_{-0.014}$ & $0.943^{+0.015}_{-0.017}$ & $0.943^{+0.017}_{-0.013}$ \\
    $\ln[10^{10}A_s]$ & $3.049^{+0.037}_{-0.037}$ & $3.050^{+0.040}_{-0.035}$ & $3.049^{+0.034}_{-0.039}$ & $3.050^{+0.032}_{-0.040}$ \\
    \hline
    age~[Gyr] & $13.92^{+0.29}_{-0.34}$ & $13.85^{+0.17}_{-0.22}$ & $13.97^{+0.16}_{-0.15}$ & $13.96^{+0.17}_{-0.15}$ \\
    $\Omega_m$ & $0.27^{+0.03}_{-0.13}$ & $0.254^{+0.030}_{-0.024}$ & $0.266^{+0.035}_{-0.034}$ & $0.264^{+0.019}_{-0.020}$ \\
    $z_\mathrm{re}$ & $10.4^{+1.4}_{-1.2}$ & $10.5^{+1.3}_{-1.3}$ & $10.4^{+1.3}_{-1.4}$ & $10.4^{+1.4}_{-1.2}$ \\
    $H_0$ & $75^{+13}_{-17}$ & $74.1^{+3.9}_{-3.2}$ & $73.7^{+4.7}_{-6.8}$ & $73.5^{+3.2}_{-3.0}$ \\
    $m_\nu$~[eV] & $<0.44 ~(95\%)$ & $<0.42 ~(95\%)$ & $<0.48 ~(95\%)$ & $<0.46 ~(95\%)$ \\
    \hline
    \hline
  \end{tabular}
  \caption{Constraints on the $w$CDM+$m_\nu$ model from current CMB
    data, combined with and without the BAO data and the direct
    measurement of $H_0$.  }
  \label{table:current_wCDM}
  \end{center}
\end{table}

\begin{table}
  \begin{center}
  \begin{tabular}{lrrrr}
    \hline
    \hline
    parameters & CMB+SHOES & CMB+BAO & CMB+BAO+SHOES \\
    \hline
    $\omega_b\times10^2$ & $2.271^{+0.050}_{-0.057}$ & $2.262^{+0.045}_{-0.061}$ & $2.271^{+0.059}_{-0.046}$   \\
    $\omega_c$ & $0.1076^{+0.0047}_{-0.0047}$ & $0.1085^{+0.0044}_{-0.0049}$ & $0.1089^{+0.0053}_{-0.0039}$  \\
    $\theta_s\times10^2$ & $104.17^{+0.23}_{-0.21}$ & $104.15^{+0.23}_{-0.21}$ & $104.17^{+0.22}_{-0.21}$ \\
    $\tau$ & $0.088^{+0.016}_{-0.017}$ & $0.088^{+0.015}_{-0.017}$ & $0.088^{+0.013}_{-0.019}$ \\
    $\Omega_k$ & $0.006^{+0.006}_{-0.010}$ & $0.000^{+0.005}_{-0.010}$ & $-0.0018^{+0.0065}_{-0.0084}$ \\
    $\omega_\nu$ & $<0.011~(95\%)$ & $<0.012~(95\%)$ & $<0.011~(95\%)$ \\
    $n_s$ & $0.957^{+0.014}_{-0.012}$ & $0.954^{+0.014}_{-0.012}$ & $0.9560^{+0.0065}_{-0.0084}$ \\
    $\ln[10^{10}A_s]$ & $3.046^{+0.036}_{-0.038}$ & $3.047^{+0.036}_{-0.038}$ & $3.0510^{+0.037}_{-0.037}$ \\
    \hline
    age[Gyr] & $13.51^{+0.33}_{-0.38}$ & $13.94^{+0.24}_{-0.32}$ & $13.78^{+0.25}_{-0.25}$  \\
    $\Omega_m$ & $0.256^{+0.024}_{-0.030}$ & $0.294^{+0.020}_{-0.024}$ & $0.2815^{+0.013}_{-0.023}$ \\
    $z_\mathrm{re}$ & $10.5^{+1.4}_{-1.7}$ & $10.5^{+1.3}_{-1.3}$ & $10.5^{+1.2}_{-1.3}$ \\
    $H_0$ & $72.8^{+3.2}_{-3.5}$ & $68.0^{+2.1}_{-2.1}$ & $69.6^{+1.9}_{-1.6}$ \\
    $m_\nu$~[eV]  & $<0.36~(95\%)$ &  $<0.37~(95\%)$ & $<0.35~(95\%)$ \\
    \hline
    \hline
  \end{tabular}
  \caption{Constraints on $\Omega\Lambda$CDM+$m_\nu$ model from
    current observations.  }
  \label{table:current_oLCDM}
  \end{center}
\end{table}

\begin{table}
  \begin{center}
  \begin{tabular}{lrrrr}
    \hline
    \hline
    parameters & CMB+SHOES & CMB+BAO & CMB+BAO+SHOES \\
    \hline
    $\omega_b\times10^2$ & $2.246^{+0.052}_{-0.060}$ & $2.241^{+0.052}_{-0.058}$ & $2.241^{+0.045}_{-0.063}$   \\
    $\omega_c$ & $0.1097^{+0.0049}_{-0.0051}$ & $0.1099^{+0.0045}_{-0.0054}$ & $0.1100^{+0.0043}_{-0.0050}$  \\
    $\theta_s\times10^2$ & $104.13^{+0.22}_{-0.22}$ & $104.12^{+0.21}_{-0.22}$ & $104.11^{+0.24}_{-0.19}$ \\
    $\tau$ & $0.086^{+0.016}_{-0.016}$ & $0.086^{+0.015}_{-0.017}$ & $0.086^{+0.014}_{-0.017}$ \\
    $\Omega_k$ & $0.009^{+0.014}_{-0.016}$ & $-0.006^{+0.006}_{-0.010}$ & $-0.007^{+0.006}_{-0.010}$ \\
    $\omega_\nu$ & $<0.013~(95\%)$ & $<0.013~(95\%)$ & $<0.013~(95\%)$ \\
    $w_X$ & --- & --- & $-1.47^{+0.29}_{-0.27}$ \\
    $n_s$ & $0.948^{+0.016}_{-0.013}$ & $0.947^{+0.016}_{-0.013}$ & $0.947^{+0.015}_{-0.013}$ \\
    $\ln[10^{10}A_s]$ & $3.046^{+0.036}_{-0.037}$ & $3.046^{+0.034}_{-0.039}$ & $3.047^{+0.032}_{-0.039}$ \\
    \hline
    age[Gyr] & $14.4^{+0.8}_{-1.0}$ & $14.32^{+0.42}_{-0.46}$ & $14.29^{+0.41}_{-0.40}$  \\
    $\Omega_m$ & $0.256^{+0.024}_{-0.031}$ & $0.250^{+0.039}_{-0.046}$ & $0.254^{+0.023}_{-0.024}$ \\
    $z_\mathrm{re}$ & $10.4^{+1.4}_{-1.3}$ & $10.4^{+1.3}_{-1.3}$ & $10.4^{+1.5}_{-1.1}$ \\
    $H_0$ & $73.5^{+3.6}_{-3.5}$ & $75.0^{+5.1}_{-7.9}$ & $73.9^{+2.7}_{-3.5}$ \\
    $m_\nu$~[eV]  & $<0.40~(95\%)$ &  $<0.42~(95\%)$ & $<0.41~(95\%)$ \\
    \hline
    \hline
  \end{tabular}
  \caption{Constraints on $\Omega w$CDM+$m_\nu$ model from current
    observations.  Note that constraints on $w_X$ from CMB+SHOES and
    CMB+BAO are not presented, since the posterior probabilities of
    $w_X$ are not bounded in the prior range, $w_X \in [-3,0]$.  }
  \label{table:current_owCDM}
  \end{center}
\end{table}

\begin{figure}[t]
\begin{center}
  \scalebox{0.75}{
  \includegraphics{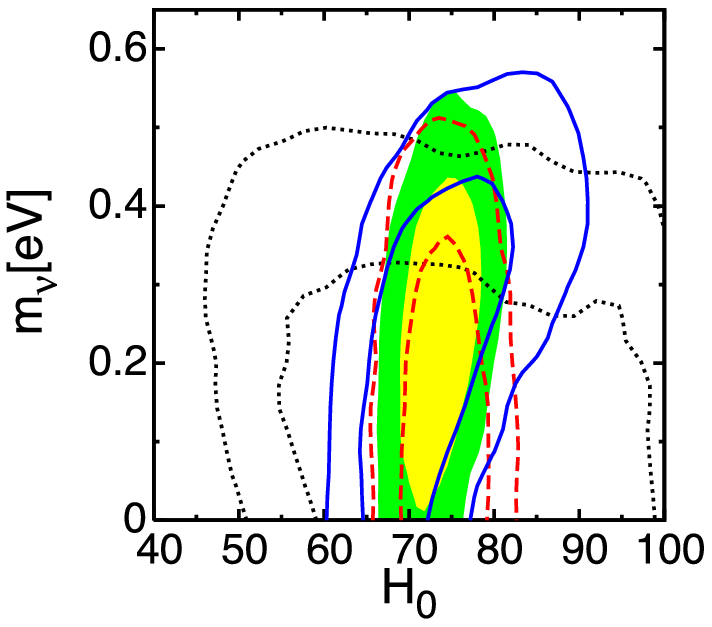}\hspace{0mm}
  \includegraphics{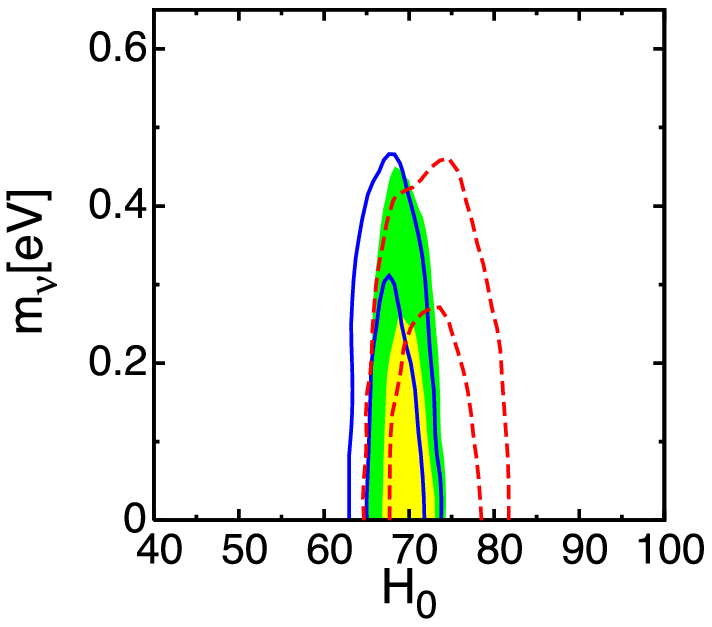}\hspace{0mm}
  \includegraphics{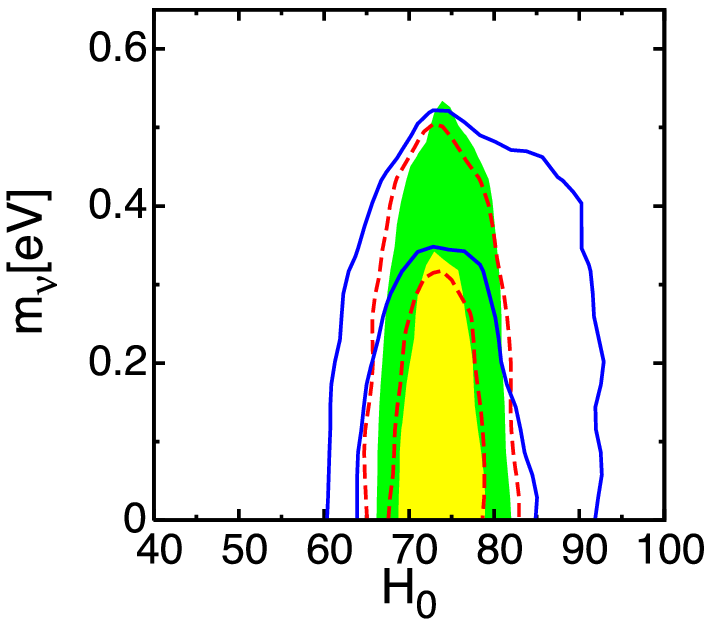}}
\caption{ Confidence regions of 68\% and 95\% C.L. from current CMB
  data on the $m_\nu$--$H_0$ plane for the $w$CDM model (left),
  $\Omega\Lambda$CDM model (middle) and $\Omega w$CDM model (right).
  Presented are the results from the present CMB data alone (black dotted
  line), CMB with SHOES prior $H_0 = 74.2 \pm3.6$ (red dashed
  line), CMB with BAO data(blue solid line), and CMB combined
  with BAO and SHOES (green and yellow shaded regions). Inner (outer)
  contours and shaded regions show regions of 68\%(95\%) C.L.
  Note that $H_0$ is not explored at $H_0>100$ nor $H_0<40$ 
  in MC analyses. Note that analyses with CMB data alone are not performed
  for the $\Omega\Lambda$CDM and $\Omega w$CDM models (See text for details).}
\label{fig:mnu_h_current_wCDM}
\end{center}
\end{figure}

Finally we investigate future constraints on neutrino masses in
$w$CDM, $\Omega \Lambda$CDM and $\Omega w$CDM model.  We repeat the
analysis using the expected data from Planck, direct $H_0$ measurement
and BAO scale as we did in Sec.~\ref{sec:planck} for these models.  We
summarized the marginalized constraints on $m_\nu$ in
Tables~\ref{table:Planck_mnu0} and \ref{table:Planck_mnu01} for
fiducial values of $m_\nu =0$ and $0.1\,\mbox{eV}$, respectively.
Hereafter we mainly discuss the constraints for the case of the
fiducial value of the neutrino mass of $m_\nu=0\,\mbox{eV}$, because
there is no qualitative difference between fiducial models of
$m_\nu=0$ and $0.1$~eV.
  
\begin{figure}[t]
\begin{center}
  \begin{tabular}{cc}
    \scalebox{0.8}{\includegraphics{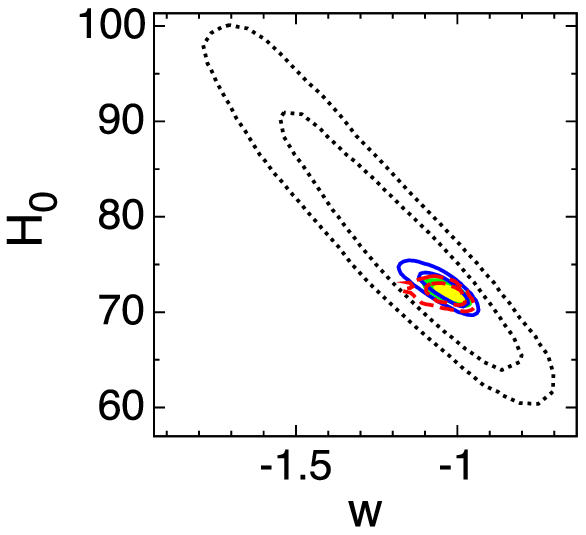}}\hspace{0mm} \\
    \scalebox{0.8}{\includegraphics{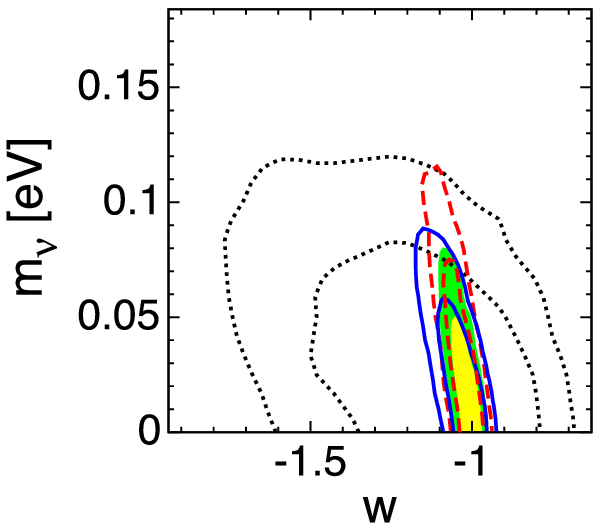}}\hspace{0mm} &
    \scalebox{0.8}{\includegraphics{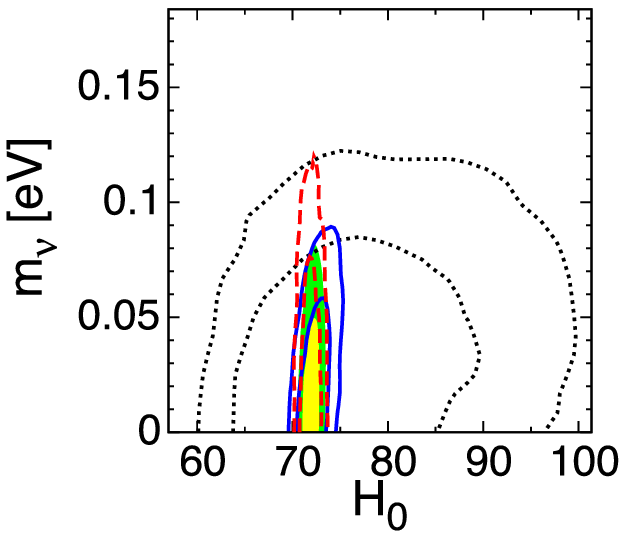}}
  \end{tabular}
  \caption{ Regions of 68\% and 95\% C.L. for the $w$CDM model using
    expected data of future observations. We assume
    $m_\nu=0\,\mbox{eV}$ as a fiducial value. Shown are constraints
    from Planck alone (black dotted line), Planck+$H_0$ (red dashed
    line), Planck+BOSS (blue solid line) and Planck+$H_0$+BOSS (green
    and yellow shaded regions). Inner (outer) contours and shaded
    regions show regions of 68\% (95\%) C.L.}
\label{fig:Planck_wCDM_2d}
\end{center}
\end{figure}

Let us first discuss the case with the $w$CDM model.  For the Planck
only analysis, $w_X$ degenerates with $H_0$ most significantly, which
can be seen in Figure~\ref{fig:Planck_wCDM_2d} where we present 2d
contours of marginalized posterior distributions among $m_\nu$, $w_X$
and $H_0$.  As mentioned for the constraints from current data, the
neutrino mass is effectively irrelevant to the degeneracy in the
geometric distance, particularly in the CMB analysis in this model.
Due to this strong degeneracy between $w_X$ and $H_0$, even if we add
the data from the $H_0$ measurement, the constraint on $m_\nu$ does
not improve much unlike the case of the $\Lambda$CDM model since the
effect of $m_\nu$ can be easily cancelled by changing $w_X$.  On the
other hand, when we add the BAO data to Planck, the constraint on
$m_\nu$ becomes slightly stronger.  Since the BAO scale measurement
can probe the geometric distance up to $z=2.5$, it is sensitive to
both $w_X$ and $H_0$. Thus the effect of varying $m_\nu$ on the
distance measure can only be partly compensated by the change of $w_X$
and $H_0$ in this case. Hence the situation is a bit different from
the case of Planck$+H_0$ where $w_X$ can be relatively freely adjusted
to cancel the effect of varying $m_\nu$. In any case, to obtain a
severe constraint on $m_\nu$, we need to determine the values of both
$H_0$ and $w_X$ accurately, which motivates us to include both
measurements of $H_0$ and the BAO scale. As seen from
Table~\ref{table:Planck_mnu0}, the constraint on $m_\nu$ from
Planck+BOSS+$H_0$ is much severer than that from Planck alone. Thus
the direct measurement of $H_0$ should also have a strong power in
constraining $m_\nu$ in combination with some other distance
measurement such as BAO in this case as well.

\begin{figure}[t]
\begin{center}
  \begin{tabular}{cc}
    \scalebox{0.8}{\includegraphics{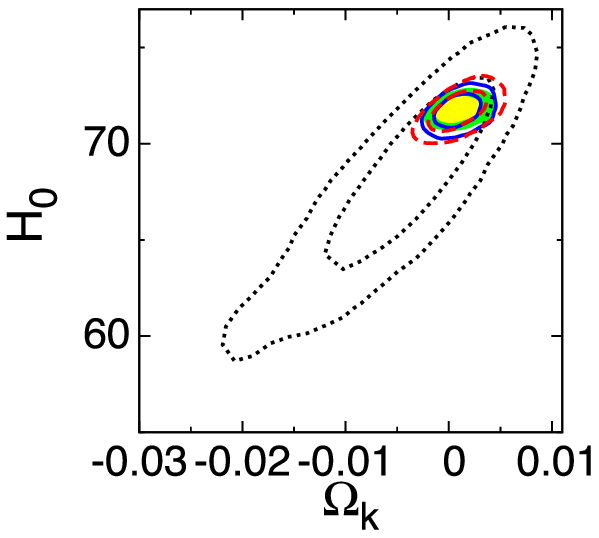}}\hspace{0mm} \\
    \scalebox{0.8}{\includegraphics{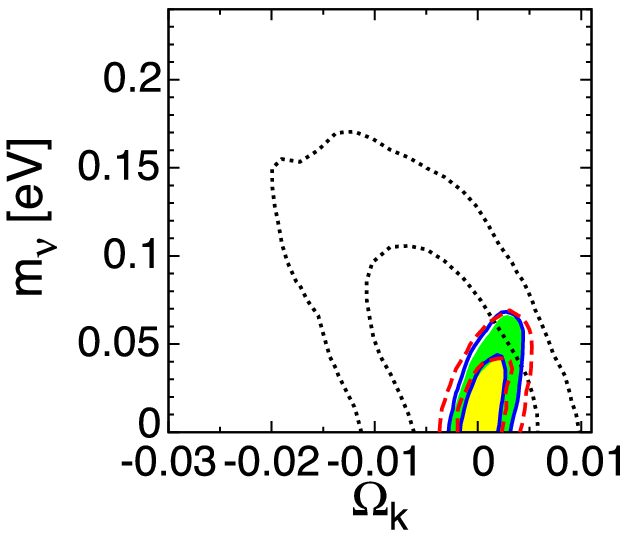}}\hspace{0mm} &
    \scalebox{0.8}{\includegraphics{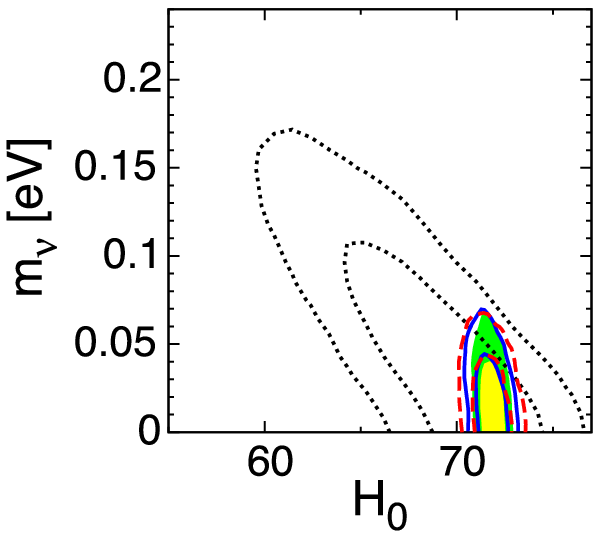}}
  \end{tabular}
  \caption{ Same as in Figure~\ref{fig:Planck_wCDM_2d} but for
    $\Omega\Lambda$CDM model.}
\label{fig:Planck_oCDM_2d}
\end{center}
\end{figure}

Next we look at the case of the $\Omega\Lambda$CDM model.  In this
case, the situation is somewhat different (See
Figure~\ref{fig:Planck_oCDM_2d}).  One of noticeable differences is
that $m_\nu$, $H_0$ and $\Omega_k$ are notably correlated with one
another in CMB.  This degeneracy shows that the change in the angular
diameter distance to last scattering surface made by varying one of
the above parameters cannot be efficiently compensated by just
adjusting another. Thus other observations of the geometrical
distances would be necessary. When the BAO scale measurement is
included, the constraint on $m_\nu$ is improved because of a better
determination of $\Omega_k$ and $H_0$.  For the analysis of
Planck+$H_0$, the constraint on $m_\nu$ also becomes stronger.  As
seen from Table~\ref{table:Planck_mnu0}, Planck+BOSS and Planck+$H_0$
give almost the same limit on $m_\nu$.  However, even if both the BAO
scale and a direct $H_0$ measurements are included in the analysis,
the constraint is not so improved compared to Planck+BOSS and
Planck+$H_0$.  This is because the BAO scale measurement effectively
gives the same information as the direct $H_0$ measurement in this
model, which is also the reason why the limit on $m_\nu$ from
Planck+BOSS and Planck+$H_0$ are almost the same level.

\begin{figure}[t]
\begin{center}
  \begin{tabular}{ccc}
    \scalebox{0.8}{\includegraphics{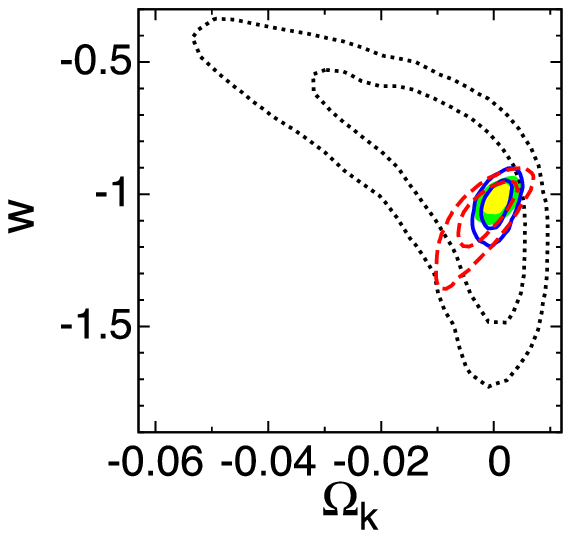}}\hspace{0mm} \\
    \scalebox{0.8}{\includegraphics{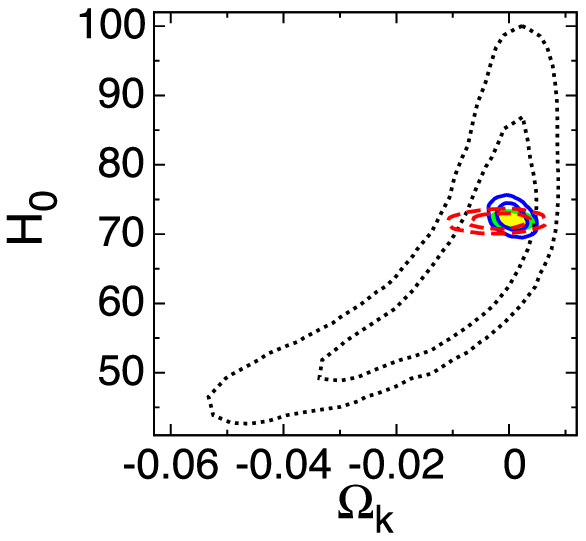}}\hspace{0mm} &
    \scalebox{0.8}{\includegraphics{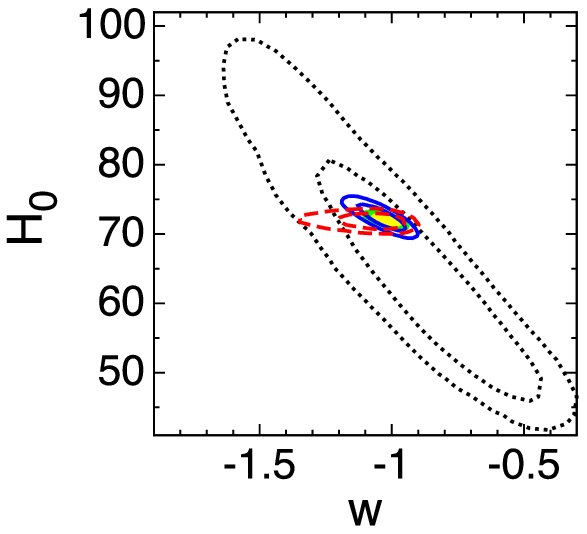}} \hspace{0mm} \\
    \scalebox{0.8}{\includegraphics{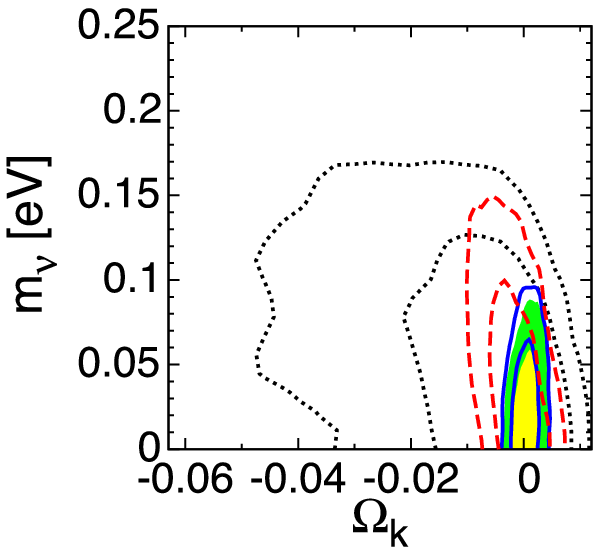}}\hspace{0mm} &
    \scalebox{0.8}{\includegraphics{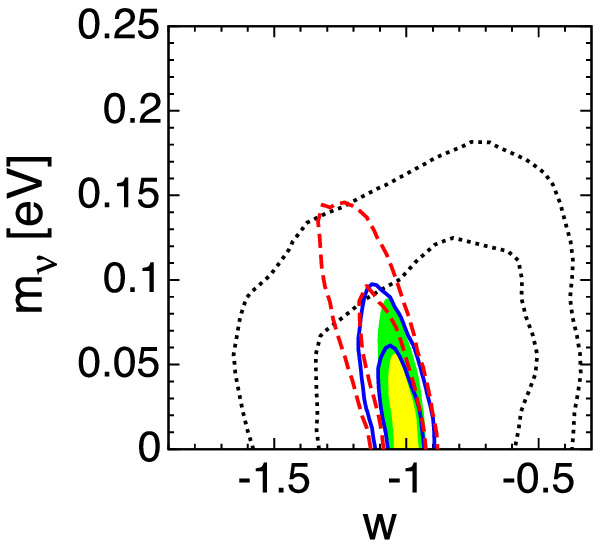}}\hspace{0mm} &
    \scalebox{0.8}{\includegraphics{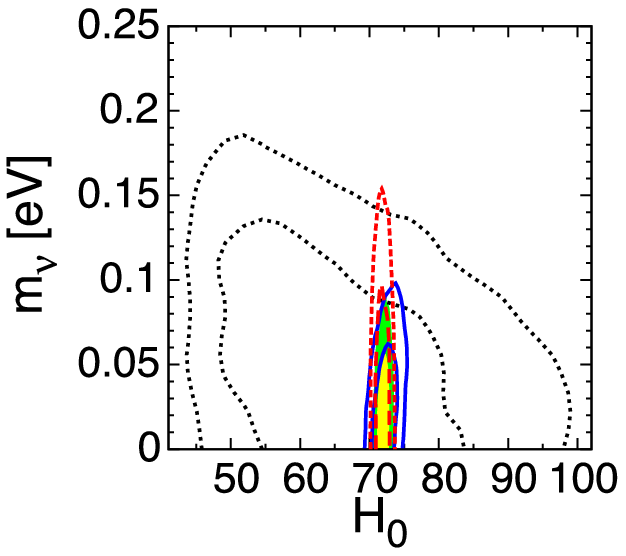}}
  \end{tabular}
  \caption{ Same as in Figure~\ref{fig:Planck_wCDM_2d} but for $\Omega w$CDM model.}
\label{fig:Planck_owCDM_2d}
\end{center}
\end{figure}

Now we move on to the case of the $\Omega w$CDM model where both
$\Omega_k$ and $w_X$ are varied.  In Fig.~\ref{fig:Planck_owCDM_2d},
1$\sigma$ and 2$\sigma$ contours are depicted, which shows the
degeneracies among the key parameters, $m_\nu$, $\Omega_k$, $w$ and
$H_0$.  Since BOSS measures the geometry at several, wide-range
redshifts, the inclusion of BOSS to Planck can remove degeneracies
among these parameters more compared to the case of Planck+$H_0$ in
this model.  However, when we combine the BOSS survey with a $H_0$
measurement at 1\% accuracy, constraints on some key parameters are
much more improved from those without data of the direct $H_0$
measurement.  Most significant improvement can be found in constraints
on $w_X$.  Even though the BAO scale measurement can measure both of
$w_X$ and $H_0$ simultaneously to some extent, there remains some
degeneracy between these parameters.  Thus adding a direct measurement
of $H_0$ would be very helpful to break the degeneracy to obtain a
better determination of $w_X$.  The improvement of constraints on
$w_X$ makes the constraint on $m_\nu$ stronger because of the residual
degeneracy between $w_X$ and $m_\nu$.

So far we have seen how the measurements of $H_0$ and the BAO scale
improve the constraints on the parameters $m_\nu$, $\Omega_k$ and
$w_X$, that are severely degenerate with one another in geometrical
distances.  The BOSS survey is now running and expected to be a
powerful tool for constraining these parameters. However, as we have
stressed, an $H_0$ measurement has its own cosmological information
independent of other observations and if a measurement at 1\,\%
accuracy is achieved, it will bring us constraints on these parameters
comparable with those from the BOSS survey, or by combining all these
together, there would be much more improvements for the constraint on
$m_\nu$.

\begin{table}
  \begin{center}
  \begin{tabular}{lrrrrr}
    \hline
    \hline
    models & Planck & Planck+$H_0$ & Planck+BOSS & Planck+BOSS+$H_0$ \\
    \hline
    $\Lambda$CDM & $<0.088~(95\%)$ & $<0.042~(95\%)$ & $<0.038~(95\%)$ & $<0.034~(95\%)$ \\
    $w$CDM & $<0.10~(95\%)$ & $<0.097~(95\%)$ & $<0.073~(95\%)$ & $<0.064~(95\%)$ \\
    $\Omega\Lambda$CDM & $<0.13$~(95\%) & $<0.055$~(95\%) & $<0.055$~(95\%) & $<0.053$~(95\%) \\
    $\Omega w$CDM & $<0.14$~(95\%) & $<0.12$~(95\%) & $<0.079$~(95\%) & $<0.070$~(95\%) \\
        \hline
    \hline
  \end{tabular}
  \caption{Summary of constraints on $m_\nu$ from future observations.
    Presented are the upper limits on $m_\nu$ at 95\% C.L. for a
    fiducial $\Lambda$CDM model with $m_\nu=0\,\mbox{eV}$.  }
  \label{table:Planck_mnu0}
  \end{center}
\end{table}

\begin{table}
  \begin{center}
  \begin{tabular}{lrrrrr}
    \hline
    \hline
    models & Planck & Planck+$H_0$ & Planck+BOSS & Planck+BOSS+$H_0$ \\
    \hline
    $\Lambda$CDM & $0.107^{+0.045}_{-0.049}$ & $0.107^{+0.021}_{-0.020}$ & $0.108^{+0.020}_{-0.019}$ & $0.107^{+0.016}_{-0.017}$ \\
    $w$CDM & $0.112^{+0.045}_{-0.051}$ & $0.110^{+0.052}_{-0.046}$ & $0.108^{+0.036}_{-0.040}$ & $0.107^{+0.029}_{-0.035}$ \\
    $\Omega\Lambda$CDM & $<0.21$~(95\%) & $0.107^{+0.024}_{-0.030}$ & $0.105^{+0.027}_{-0.027}$ & $0.105^{+0.026}_{-0.026}$ \\
    $\Omega w$CDM & $<0.21$~(95\%) & $0.112^{+0.050}_{-0.054}$ & $0.108^{+0.038}_{-0.039}$ & $0.107^{+0.032}_{-0.038}$ \\
    \hline
    \hline
  \end{tabular}
  \caption{Same table as in Table~\ref{table:Planck_mnu01}, but
    presented here are the mean values and 68\% C.L. errors for a
    fiducial $\Lambda$CDM model with $m_\nu=0.1$ eV.  }
  \label{table:Planck_mnu01}
  \end{center}
\end{table}

\section{Summary} \label{sec:result}

In this paper, we have investigated constraint of neutrino mass from
cosmological data with particular attention to direct measurements of
the Hubble constant and BAO scale as supplements to CMB data.  Since a
strong degeneracy exists between $m_\nu$ and $H_0$ in the CMB
analysis, which originates from the distance scale, neutrino mass is
not effectively constrained from the current CMB observations alone.
However, available measurement of $H_0$ breaks this degeneracy and
provides a mass constraint of $m_\nu<0.20~\mbox{eV}$ (95\% C.L.).  We
have also examined the constraint using the BAO data recently
released.  When the measurement of BAO scales is combined with the
present CMB data, we obtain a similar limit for $m_\nu$; current
Hubble constant and BAO measurements are competitive in the context of
$\Lambda$CDM models, where we emphasize the EOS of dark energy is
constrained to be $-1$ and curvature is zero.

We have also studied attainable limits on $m_\nu$ that may be
anticipated from next-generation data sets.  For this purpose, we have used
projected Planck data in combination with a future precise measurement
of $H_0$ and of the BAO scale as may be provided by BOSS before 2015.
If the Hubble constant is directly measured with precision $\Delta
H_0/H_0=0.01$, then neutrino mass can be detected at the $5\sigma$
level for a $0.1~\mbox{eV}$ mass.  Similar accuracy can be achieved
through combination of BAO and CMB data.  The uncertainty of the
Hubble constant degrades the constraint on the neutrino mass
approximately as $(\Delta H_0/H_0)^{0.4}$. A more modest accuracy of
e.g., 3\,\% does not provide a useful mass constraint and favors BAO
techniques.

In models where $w_X$ or $\Omega_k$ are allowed to vary (including
more parameters), the geometric degeneracy between $m_\nu$ and $H_0$
is broadened, and combinations of data and priors (i.e., CMB+$H_0$ or
CMB+BAO) do not strongly constrain $m_\nu$.  However when both $H_0$
measurement and BAO are combined with CMB, the constraint becomes
tighter.  In this respect, a direct measurement of $H_0$ should be
very beneficial along with other cosmological observations.

We have investigated the impact of precision measurement of $H_0$ on
the neutrino mass.  The analysis demonstrates that the application of
cosmological data has a unique contribution to make in the study of
fundamental physics.  Analyses similar to ours would plausibly
demonstrate strong constraint of other parameters, e.g., $w_X$
\cite{Ichikawa:2007je,Greenhill:2009yi}, though degeneracies will vary
case to case.  In general cases (e.g., nonzero curvature and $w_X \ne
-1$), the constraints imposed by future BAO and $H_0$ measurements, if
of comparable accuracy, are likely contribute at about the same level,
but the larger number of model parameters argues pursuit of both BAO
and $H_0$ measurements, so as to leverage the independence of
systematic errors in each.

\bigskip
\bigskip

\noindent 
\section*{Acknowledgment}
We are very grateful to Ingyin Zaw and Marc Anera for useful comments
on the draft.  T.S. would like to thank the Japan Society for the
Promotion of Science for financial support.  This work is partially
supported by Grant-in-Aid for Scientific research from the Ministry of
Education, Science, Sports, and Culture, Japan, No. 19740145 (T.T.).


\end{document}